\documentclass[%
preprint,
 amsmath,amssymb,
 aps,
]{revtex4-2}

\usepackage{graphicx}
\usepackage{dcolumn}
\usepackage{bm}

\usepackage{ulem} 
\usepackage{color}
\usepackage{natbib}
\setcitestyle{authoryear,open={(},close={)}} 

\newcommand{\commentout}[1]{}

\newcommand{\bs}{\boldsymbol}

\newcommand{\iimag}{\mathrm{i}}
\newcommand{\ba}{{\bs{a}}}
\newcommand{\bA}{{\bs{A}}}
\def\bb {\bs{b}}
\def\bB {\bs{B}}
\def\bn {{\bs{n}}}
\def\br {\bs{r}}
\def\bW {\bs{W}}

\newcommand{\bv}{{\bf v}}
\newcommand{\bu}{{\bf u}}

\newcommand{\bk}{{\bf k}}
\newcommand{\bF}{{\bf f}}
\newcommand{\bp}{{\bf p}}
\newcommand{\bq}{{\bf q}}
\newcommand{\bx}{{\bf x}}
\newcommand{\half}{\frac{1}{2}}
\newcommand{\bD}{{\bf D}}
\newcommand{\bG}{{\bf G}}
\newcommand{\bL}{{\bf L}}
\newcommand{\btau}{{\mbox{\boldmath $\tau$}}}

\newcommand{\be}{\begin{equation}}
\newcommand{\ee}{\end{equation}}
\newcommand{\bea}{\begin{eqnarray}}
\newcommand{\eea}{\end{eqnarray}}
\newcommand{\lb}{\label}

\newcommand{\grad}{{\mbox{\boldmath $\nabla$}}}
\newcommand{\bdot}{{\mbox{\boldmath $\cdot$}}}

\newcommand{\bzed}{{\mbox{\boldmath $0$}}}

\newtheorem{Prop}{Proposition} 

\newcommand{\etal}{{\it et al.}}

\def\btau {\boldsymbol{\tau}}
\newcommand{\StressTensor}{\bs{\tau}}

\def\kB {\mathrm{k_B}}              
\def\viscKin {\nu}                  
\def\viscDyn {\mu}                  
\def\EnergyDis {\langle\epsilon\rangle}     
\def\KolLen {\eta}               
\def\KolK {k_{\eta}}                    
\def\KolVel {u_{\eta}}                  
\def\KolTemp {\theta_\eta}            
\def\KolTime {\tau_{\eta}}             
\def\kCross {k_{\theta}}
\def\MicroLen {\lambda_\mathrm{mic}}              
\def\KE {\mathcal{E}}                       
\def\ReNum {\mathrm{Re}}                    
\def\TayRe {\mathrm{Re}_\mathrm{\lambda}}   
\def\TurbRe {\mathrm{Re}_\mathrm{T}}        
\def\Skew {\mathcal{S}}             
\def\Kurt {\mathcal{K}}             

\def\conoise {\bs{\mathcal{W}}}
\def\fonoise {\bs{{W}}}
\def\dconoise {\bs{\mathcal{Z}}}
\def\dfonoise {\bs{{Z}}}

\begin{document}

\preprint{PREPRINT}

\title{Thermal Fluctuations in the Dissipation Range \\of Homogeneous Isotropic Turbulence}

\author{John B. Bell}
\email{jbbell@lbl.gov}
\author{Andrew Nonaka}%
\affiliation{Center for Computational Sciences and Engineering, Lawrence Berkeley National Laboratory
}%

\author{Alejandro L. Garcia}
\affiliation{
Dept. Physics \& Astronomy, San Jose State University}%

\author{Gregory Eyink}
\affiliation{Dept. Applied Mathematics \& Statistics, Johns Hopkins University}

\date{\today}

\begin{abstract}
Using fluctuating hydrodynamics we investigate the effect of thermal fluctuations in the dissipation range of 
homogeneous, isotropic turbulence. Simulations confirm theoretical predictions that the energy spectrum is dominated by these fluctuations at length scales comparable to the Kolmogorov length. 
We also find that the extreme intermittency 
in the far-dissipation range predicted by Kraichnan is replaced by Gaussian thermal equipartition. 
\end{abstract}

\maketitle


\section{Introduction}

At macroscopic scales, fluid dynamics is governed by partial differential equations
that characterize the behavior of a fluid in terms of fields
that represent density, momentum, and other locally conserved quantities.  
The evolution of these fields is given by fluxes that are modeled at the macroscale as
smooth, continuous functions of space and time.
However, at atomic scales fluids are discrete systems composed of individual molecules with complex interactions resulting in stochastic dynamics (e.g., Brownian motion).
These two descriptions overlap at the mesoscale.
While it is possible to describe a fluid using macroscopic field variables at the mesoscale,
the fluxes are no longer smooth; instead they include spontaneous thermal fluctuations even for systems
that are at thermodynamic equilibrium.

Turbulent fluctuations are of a completely different origin. They are produced from the nonlinear cascade of energy due to external forcing at large length scales down to the scale of viscous dissipation.  
Below the length scale of the external forcing
the energy spectrum of turbulent fluctuations has two regimes, inertial and dissipative.
The demarcation between them is determined by the Kolmogorov length scale $\KolLen \equiv (\viscKin^3/\EnergyDis)^{1/4}$ where $\viscKin$ is the kinematic viscosity and $\EnergyDis$ is mean energy dissipation rate. In the inertial range (wavenumber $k \ll \KolK \equiv 2\pi/\KolLen$) the energy spectrum has the form \citep{Kolmogorov1941local,BatchelorBook1953}, $E(k) \propto k^{-5/3}$, 
while in the dissipative range it is often modeled as,
$E(k) \propto k^\alpha \exp( - \beta k )$
\citep{Frisch1995,PopeTurbulentFlows2001,Buaria2020PRF}.
Since $E(k)$ decreases exponentially in the dissipative range it is natural to ask:
At what wavenumber do mesoscopic thermal fluctuations have a significant effect on this spectrum?
This question dates back to the pioneering work of Betchov \citep{Betchov1957,Betchov1961} and has received renewed attention in recent theoretical studies \citep{Eyink2021,bandak2021} and molecular simulation work \citep{Gallis2021}. 

To investigate the influence of thermal fluctuations on turbulence we turn to the theory of fluctuating hydrodynamics (FHD), originally proposed by Landau and Lifshitz \citep{Landau1959Fluid}.
This theory extends conventional hydrodynamics by augmenting each dissipative flux with a random field.
Over the past decade our group and others have extended FHD, combining the derivation of models for complex fluids at the mesoscale with the development of numerical algorithms for solving
the resulting systems on high-performance computers.
For example, we developed low Mach number stochastic FHD models for isothermal, non-ideal liquid mixtures 
that exploit the separation of scales between fluid motion and acoustic wave propagation \citep{Nonaka2015Camcos}.  
It has been demonstrated that this methodology can accurately model a wide range of phenomena both near and far from equilibrium (e.g., the experimentally observed ``giant fluctuations'' effect in fluid mixing \citep{DonevPRL2011}).

In this paper we use fluctuating hydrodynamics to investigate the effect of thermal fluctuations in the dissipation range for homogeneous, isotropic turbulence.
Section~\ref{sec:Theory} summarizes the formulation of incompressible FHD and estimates the
wavenumber at which thermal fluctuations dominate turbulent fluctuations in the energy spectrum.
Section~\ref{sec:Numerics} outlines the numerical algorithm of the simulations and
Section~\ref{sec:Results} presents their results.
In brief, the simulations confirm that thermal fluctuations dominate the energy spectrum at 
length scales comparable to the Kolmogorov length and produce nearly Gaussian velocity statistics 
at somewhat smaller scales. 

\section{Fluctuating hydrodynamics with forcing}
\label{sec:Theory}

In fluctuating hydrodynamics we write the incompressible, isothermal stochastic Navier-Stokes equations as,
\begin{align}
    \partial_t \; (\rho \bu ) &= -\nabla\cdot(\rho\bu\bu^T) - \nabla \pi  -\nabla\cdot(\bar{\StressTensor}+\tilde{\StressTensor}) + \rho\, \ba^F(\br,t), \nonumber \\
    \nabla \cdot \bu &= 0,
    \label{eq:FNS}
\end{align}
where $\bu$ is the velocity, $\rho$ is the density, $\pi$ is a perturbational pressure that enforces the divergence constraint and $\ba^F$ is a long wavelength acceleration.
Here the deterministic stress tensor is $\bar{\StressTensor} = -\viscDyn[\nabla \bu + (\nabla \bu)^T]$ with dynamic viscosity $\viscDyn$. The stochastic stress tensor chosen according to the fluctuation-dissipation relation \citep{Landau1959Fluid,ZarateBook2006} is
\begin{align}
    \tilde{\StressTensor} = \sqrt{\viscDyn k_B T}~(
    \conoise + \conoise^T
    ),
\end{align}
where $\conoise$
is a standard white noise Gaussian tensor with uncorrelated components.

The long wavelength acceleration, $\ba^F$, due to an external
force added to drive turbulence, is modeled here using the formulation of \citep{Eswaran1988CompFluids}. 
Define a Ornstein-Uhlenbeck (OU) process for the complex vector $\bb(\bn , t)$ as,
\begin{align}
    d\bb(\bn) = \bA \bb(\bn) ~dt + \bB ~d\fonoise, 
    \label{eq:UO}
\end{align}
where $\bn = (n_x, n_y, n_z)$ are integer indexes 
such that $1 \leq | \bn | \leq n_\mathrm{max}$ 
and $\fonoise$ is a vector of complex Wiener processes. 
The external forcing is limited to long wavelengths by taking $n_\mathrm{max} = 2\sqrt{2}$.  
The matrices in the OU process are taken to be,
\begin{align}
    \bA = \frac{1}{T_L} \bs{I} \qquad ; \qquad 
    \bB = \sigma \sqrt{\frac{1}{T_L}}\bs{I},
\end{align}
where $\bs{I}$ is the identity matrix.
In this case we have \citep{Gardiner1985book},
\begin{align}
    \langle \bb(\bn,t) , \bb^*({\bn'},t + s) \rangle = \frac{\sigma^2}{2} e^{- s/T_L}  \delta_{\bn,\bn'},
\end{align}
so the parameters $T_L$ and $\sigma$ are the characteristic time scale and amplitude of the
acceleration due to the external forcing.
We then define the forcing in real space as
\begin{equation}
\ba^F (\br,t) = \Re \left[ \sum_{|\bn| \leq n_\mathrm{max}}   (\bb(\bn) + \bb^*({-\bn}) )~ e^{\iimag \; \bk \cdot \br} \right ],
\label{eq:forcing}
\end{equation}
where $\bk = 2\pi \bn/L$, $L$ is the domain length,
and $\Re$ denotes the real part of a complex number. 
Note that the enforcement of the divergence free constraint on
$\ba^F$ is handled automatically by the solution algorithm. 

For a given velocity field, $\bu(\br)$, the Fourier transform is defined as,
\begin{equation}
    \hat{\bu}(\bk) =
    \int  d\br ~e^{\iimag \bk \cdot \br} \bu( \br ),
\end{equation}
The specific energy density is $E(\bk) = \frac{1}{2}\langle\hat{u}(\bk)\cdot\hat{u}^*(\bk)\rangle$,
its isotropic integral is $E(k)$,
and the total turbulent kinetic energy is $\KE = \int d\bk E(\bk)$. 
The specific dissipation rate is
\begin{equation}
   \EnergyDis  = \viscKin \left\langle \frac{\partial u_i}{\partial x_j}\frac{\partial u_i}{\partial x_j} \right\rangle
   = - \viscKin \langle u_j \nabla^2 u_j \rangle = 2 \viscKin \int d\bk k^2 E(\bk),
\end{equation}
 where $\viscKin = \viscDyn/\rho$ is the kinematic viscosity. 
 
In the absence of forcing the deterministic Navier-Stokes equations can be characterized in terms of a single parameter, namely, the Reynolds number, 
$\ReNum = {U L}/{\viscKin}$,
where $U$ and $L$ are characteristic velocities and length scales, respectively.  (In the present setting, we can consider the forcing to simply be an, albeit indirect, approach to setting the velocity scale.)
It is also useful to define the turbulence Reynolds number, $\TurbRe = \KE^2/(\EnergyDis\viscKin)$ 
and the Taylor-scale Reynolds number, $\TayRe = (\frac{20}{3} \TurbRe)^{1/2}$. 

Without forcing the energy spectrum at long times for Eq.~(\ref{eq:FNS}) is dominated 
by the effect of thermal fluctuations.  
These fluctuations can be characterized by the covariance of the velocity field at equilibrium, which is typically referred
to as the velocity structure factor. The fluctuation-dissipation relation for the stochastic dynamics \eqref{eq:FNS} implies, in agreement with Einstein-Boltzmann theory, that
\begin{equation}
    S_{\bu,\bu} = < \widehat{(\delta \bu)} \, \widehat{(\delta \bu)}^*> = \frac{\kB T}{ \rho} \left ( \bs{I} - \frac{\bk \bk^T}{|k|^2} \right ).
    \label{eq:SF}
\end{equation}
The contribution of fluctuations to the energy spectrum is then given by
\begin{equation}
E^{\mathrm{fluc}}(\bk) = \frac{1}{2} < \widehat{(\delta \bu)}^*  \widehat{(\delta \bu)}> 
= \frac{1}{2} \mathrm{Tr} \; ( S_{\bu,\bu} )= \frac{\kB T}{ \rho} \; .
\label{eq:fluc_spec}
\end{equation}
We note that this contribution is independent of $\bk$.
Consequently, in the energy spectrum
fluctuations scale like $E(k) \propto k^2$, reflecting the scaling of surface area with radius.  We note that the 
spatial discretization employed in this work is constructed so that this result remains exactly true 
in continuous time
(\citep{DFDB} and Appendix~A) and with any discrete time-step $\Delta t$ for the linearized dynamics 
\citep{LLNS_Staggered}.


The presence of thermal fluctuations introduces an additional scale into the problem.
This additional dependence can be characterized in terms of a dimensionless temperature
\begin{equation}
\KolTemp = \frac{\kB T}{\rho \KolVel^2 \KolLen^3} 
= \frac{\kB T \EnergyDis^\frac{1}{4}}{ \rho \nu^\frac{11}{4}},
\end{equation}
where $\KolVel = (\EnergyDis \viscKin)^{1/4}$ is the Kolmogorov velocity.
An order of magnitude estimate for the crossover wavenumber, $\kCross$, between the turbulence spectrum and the thermal fluctuation spectrum is given by \citep{Eyink2021,bandak2021},
\begin{equation}
    \KolVel^2 \KolLen \exp( -\kCross \KolLen ) \approx 
    \frac{\kB T}{\rho}~\kCross^2
\end{equation}
or
\begin{equation}
    \KolTemp \approx  \exp( -\kCross \KolLen )/(\kCross \KolLen)^{2}.
    \label{eq:kCross}
\end{equation}
For example, for $\KolTemp = 10^{-9}$ the crossover wavenumber
is $\kCross \approx 15 / \KolLen \approx 2.4 \KolK$ and for $\KolTemp = 10^{-6}$ it is $\kCross \approx 9.4 / \KolLen \approx 1.5 \KolK$.
Given that $\KolTemp \approx 10^{-7}-10^{-8}$ in the atmospheric boundary layer and in laboratory experiments \citep{Debue2018} this result predicts that thermal fluctuations are significant at scales comparable to the Kolmogorov length scale.

The continuum description of fluid transport, either macroscopic or mesoscopic, is not accurate at molecular scales \citep{Corrsin1959, Moser2006}.
This breakdown occurs below the microscopic transport length scale, $\MicroLen$; its ratio to the Kolmogorov length scale is
\begin{equation}
    \frac{\MicroLen}{\KolLen} = C \frac{\mathrm{Ma}}{\TurbRe^{1/4}},
\end{equation}
where $\mathrm{Ma}$ is the Mach number and $C$ is a constant of order one. In dilute gases $\MicroLen$ is the mean free path between collisions and in liquids it is the intermolecular spacing.
This result shows that $\KolLen \gg \MicroLen$ for subsonic, high Reynolds number turbulence.
Since laboratory experiments and molecular simulations have shown that Eqns.~(\ref{eq:FNS}) are accurate down to scales comparable to $\MicroLen$ \citep{BoonYip1991,ZarateBook2006,Donev2010Camcos,DonevPRL2011}
we use them in our numerical simulations to validate the crossover wavenumber predicted by Eq.~(\ref{eq:kCross}).

\section{Simulation method}
\label{sec:Numerics}

The fluctuating hydrodynamic simulations in this paper use the numerical method described in \citep{Nonaka2015Camcos}. 
We use a staggered-grid formulation where normal velocities are stored on the normal faces of grid cells.
We begin the time step with $\bu^n$.
We evaluate a Stokes predictor to solve for a preliminary time-advanced velocity, $\bu^{\star,n+1}$,
\begin{eqnarray}
\rho\frac{\bu^{\star,n+1}-\bu^n}{\Delta t} + \nabla\pi^{\star,n+1} &=& -\nabla\cdot(\rho\bu\bu)^n - \frac{1}{2}\nabla\cdot\bar{\btau}^n - \frac{1}{2}\nabla\cdot\bar{\btau}^{\star,n+1},\nonumber\\
&&+ \nabla\cdot\left(\sqrt{\frac{\viscDyn k_B T}{\Delta t\Delta V}}\overline\dconoise^n\right) + \ba^{F,n}, \\
\nabla\cdot\bu^{*,n+1} &=& 0,
\end{eqnarray}
where 
$\Delta V$ is the cell volume, and $\overline\dconoise = \dconoise+\dconoise^T$ where $\dconoise$ is a
tensor containing unit-variance, mean-zero, independent Gaussian random variables.

We then evaluate a Stokes corrector to solve for the time-advanced velocity, $\bu^{n+1}$,
\begin{eqnarray}
\rho\frac{\bu^{n+1}-\bu^n}{\Delta t} + \nabla\pi^{n+1} &=& -\frac{1}{2}\nabla\cdot(\rho\bu\bu)^n - \frac{1}{2}\nabla\cdot(\rho\bu\bu)^{\star,n+1} - \frac{1}{2}\nabla\cdot\bar{\btau}^n - \frac{1}{2}\nabla\cdot\bar{\btau}^{n+1}\nonumber\\
&&+ \nabla\cdot\left(\sqrt{\frac{\viscDyn k_B T}{\Delta t\Delta V}}\overline\dconoise^n\right)+ \ba^{F,n}, \\
\nabla\cdot\bu^{n+1} &=& 0.
\end{eqnarray}

The only significant difference from this earlier work is the addition of the external forcing described in the previous section.
For this external forcing calculation, Eq.~(\ref{eq:UO}), we use a simple Euler-Maruyama method; a higher accuracy integrator is not needed in this context since we are generating a random forcing.
Starting with the initial condition $\bb^0 = (0, 0, 0)^T$, 
we advance  Eq.~(\ref{eq:UO}) using
\begin{align}
    \bb^{n+1} = \bb^n + \bA \bb^n ~\Delta t + \bB \sqrt{\Delta t}~\dfonoise^n\bs{I},
    \end{align}
where we only advance $\bb(\bn)$ for values of $\bn$ needed to compute $\ba^F$.
Here
$\dfonoise$ 
is a complex Gaussian (normal) distributed random number.
We can then evaluate $\ba^{F,n}$ directly from $\bb^n$ using Eq.~(\ref{eq:forcing}).

\section{Simulation Results}
\label{sec:Results}

We ran simulations with $\rho = 1.0~\mathrm{g/cm}^3$, $T = 300$~K, $\viscDyn=0.02$ poise, which corresponds to a water-glycerol mixture \citep{Segur1951viscosity}, in a (5.12~cm)$^3$ domain with a $512^3$ grid using a time step $\Delta t = 10^{-4}$~s.  
We considered two cases. 
In each case, we ran the simulation
until it became statistically stationary ($\approx 30,000$ steps). We then restarted the simulations with and without fluctuations and ran for several thousand time steps.

Case~1 has strong forcing ($\sigma^2 = 2.5~\mathrm{cm}^2/\mathrm{s}$) and Case~2 has weak forcing ($\sigma^2 = 0.1~\mathrm{cm}^2/\mathrm{s}$); in both cases the forcing time scale is $T_L = 0.1$~s. 
In Case~1 the mean energy dissipation rate $\EnergyDis \approx 4.1~\mathrm{cm}^2/\mathrm{s}^3$ resulting in Kolmogorov length and time scales of $\KolLen = 0.037$~cm and $\KolTime = (\viscKin/\EnergyDis)^{1/2} = 0.07$~s, respectively, which corresponds to $\TurbRe \approx 3600$ and $\TayRe \approx 143$.
For the Case~2 simulations (weak forcing), the mean energy dissipation $\EnergyDis \approx 0.19~\mathrm{cm}^2/\mathrm{s}^3$ resulting in Kolmogorov length and time scales of $\KolLen = 0.081$~cm and $\KolTime = 0.32$~s, respectively, corresponding to $\TurbRe \approx 890$ and $\TayRe \approx 77$.
By comparison, $\EnergyDis$ ranges from 400 to 156,000~$\mathrm{cm}^2/\mathrm{s}^3$ in the water-glycerol experiments of \citep{Debue2018}, while in geophysical flows 
$\EnergyDis$ ranges from around
0.4~$\mathrm{cm}^2/\mathrm{s}^3$ in the upper ocean mixing layer~\citep{Thorpe2007} to 
400~$\mathrm{cm}^2/\mathrm{s}^3$ for the atmospheric boundary layer~\citep{Garratt1994}.
In our Case 1 simulations the dimensionless temperature is $\KolTemp = 2.8 \times 10^{-9}$; from Eq.~(\ref{eq:kCross}) the predicted crossover wavenumber is $\kCross \approx 2.3~\KolK$.
The dimensionless temperature for Case 2 is $\KolTemp = 1.3 \times 10^{-9}$ and $\kCross 
\approx 2.4~\KolK$ is the predicted crossover wavenumber.

\begin{figure}
    \centering
    \includegraphics[width=.8\textwidth]{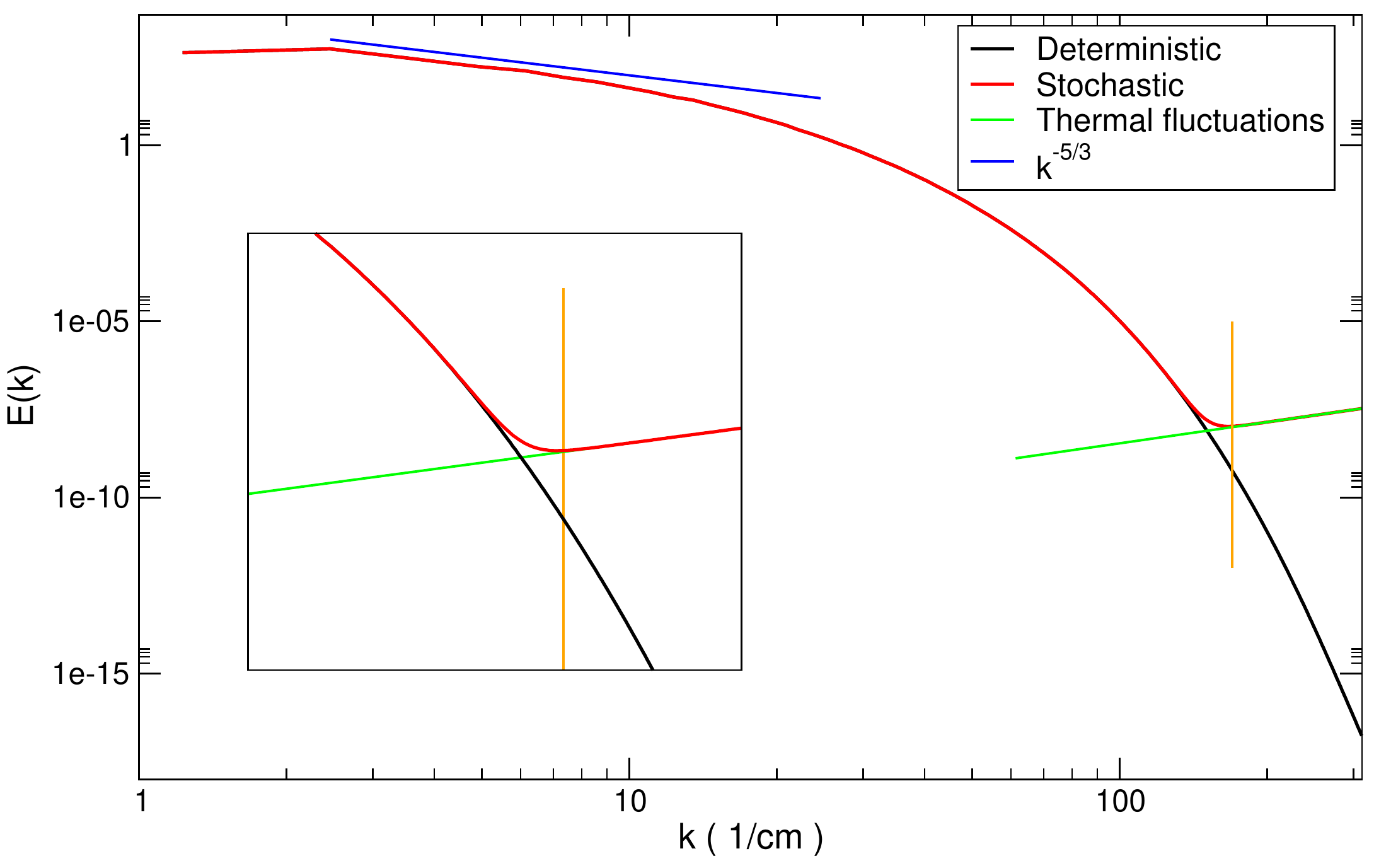}
    \caption{Specific energy density spectrum for Case 1 with thermal noise (red) and without (black); inset is region near the crossover. Blue line is $-5/3$ slope; green line represents the spectrum of thermal fluctuations given by Eq.~(\ref{eq:fluc_spec}). The Kolmogorov wavenumber $\KolK = 2\pi/\KolLen \approx 170~\mathrm{cm}^{-1}$ is indicated by the vertical orange line. 
    } 
    \label{fig:Spec_1}
\end{figure}

\begin{figure}
    \centering
    \includegraphics[width=.8\textwidth]{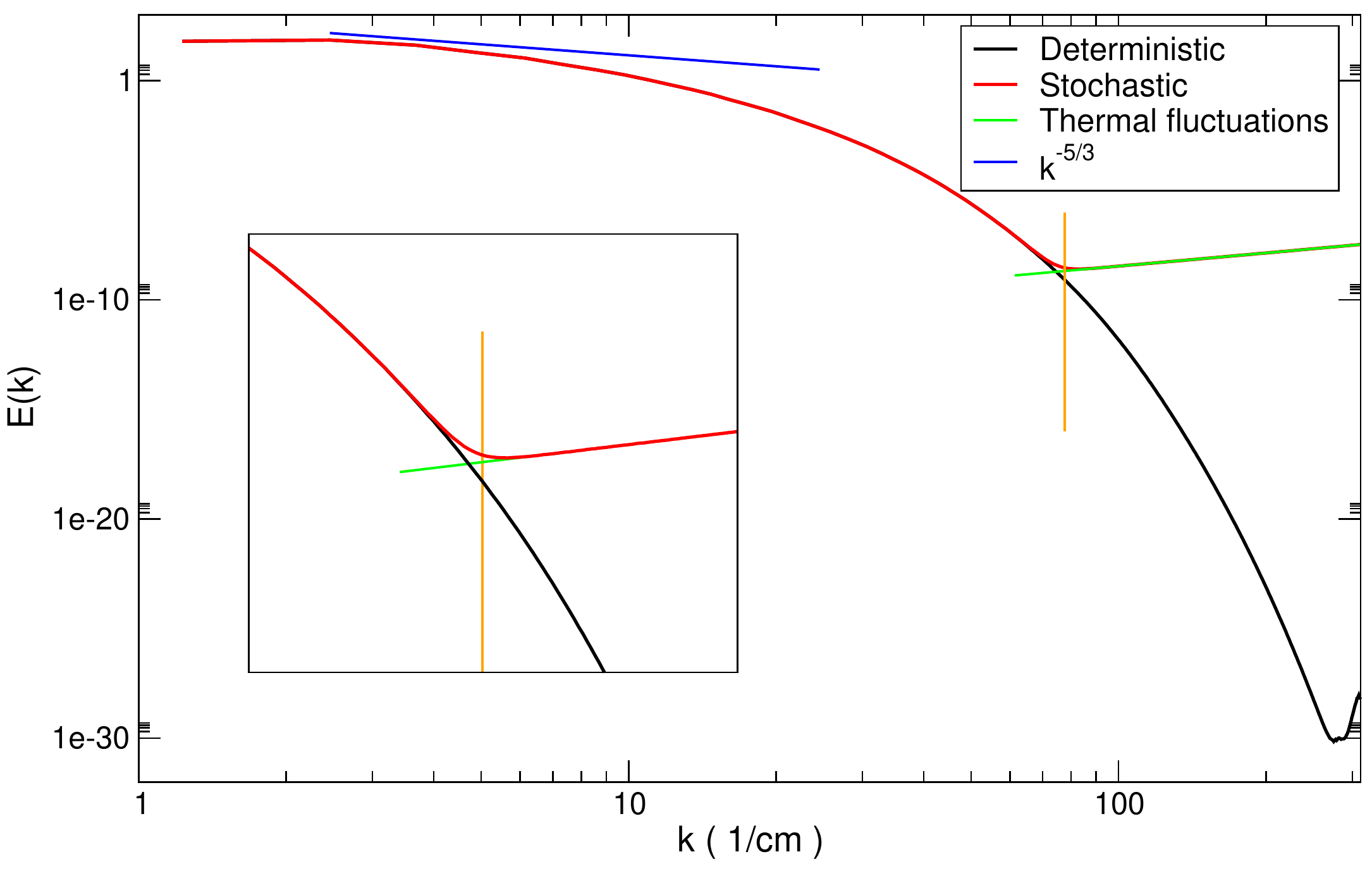}
    \caption{Specific energy density spectrum for Case 2 (see Figure~\ref{fig:Spec_1} caption); here $\KolK \approx 78~\mathrm{cm}^{-1}$. 
    }
    \label{fig:Spec_2}
\end{figure}

Spectra from Case~1 simulations, with and without thermal noise, roughly $18\KolTime$ after restart are shown in Figure \ref{fig:Spec_1} (compare with Fig.~1 in \citep{Eyink2021}). We emphasize that the green line in the figure corresponds to the theoretical spectrum of the fluctuations given by Eq.~(\ref{eq:fluc_spec}).
Spectra from Case-2 simulations, with and without thermal noise, roughly $4\KolTime$ after 
restart are shown in Figure \ref{fig:Spec_2}. 
The crossover between the turbulence spectrum and the thermal fluctuation spectrum occurs at $k \approx \KolK$, in agreement with the order of magnitude estimate given by $\kCross$. 

\begin{figure}
    \centering
    \includegraphics[width=.7\textwidth]{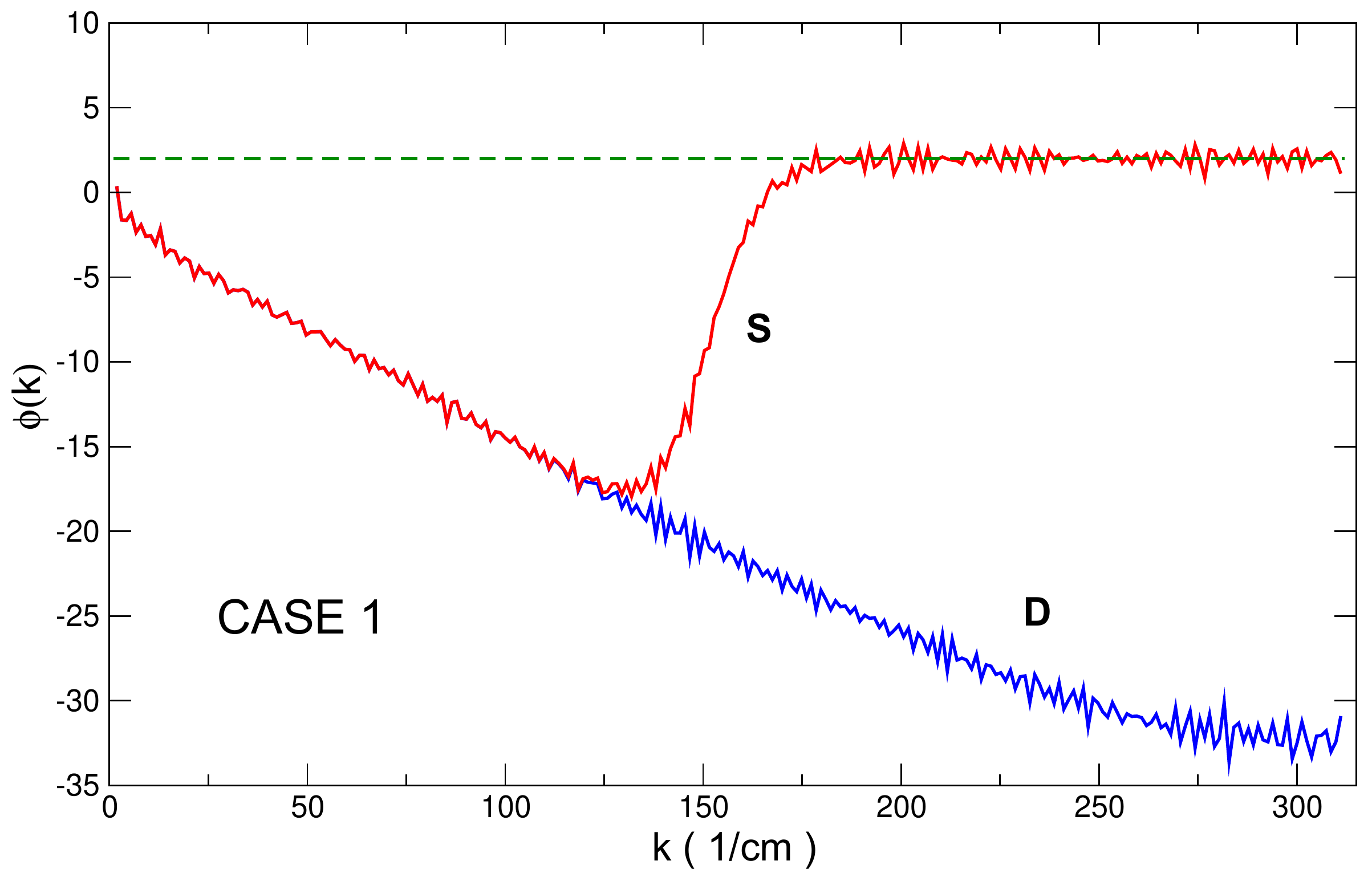} \\
     \includegraphics[width=.7\textwidth]{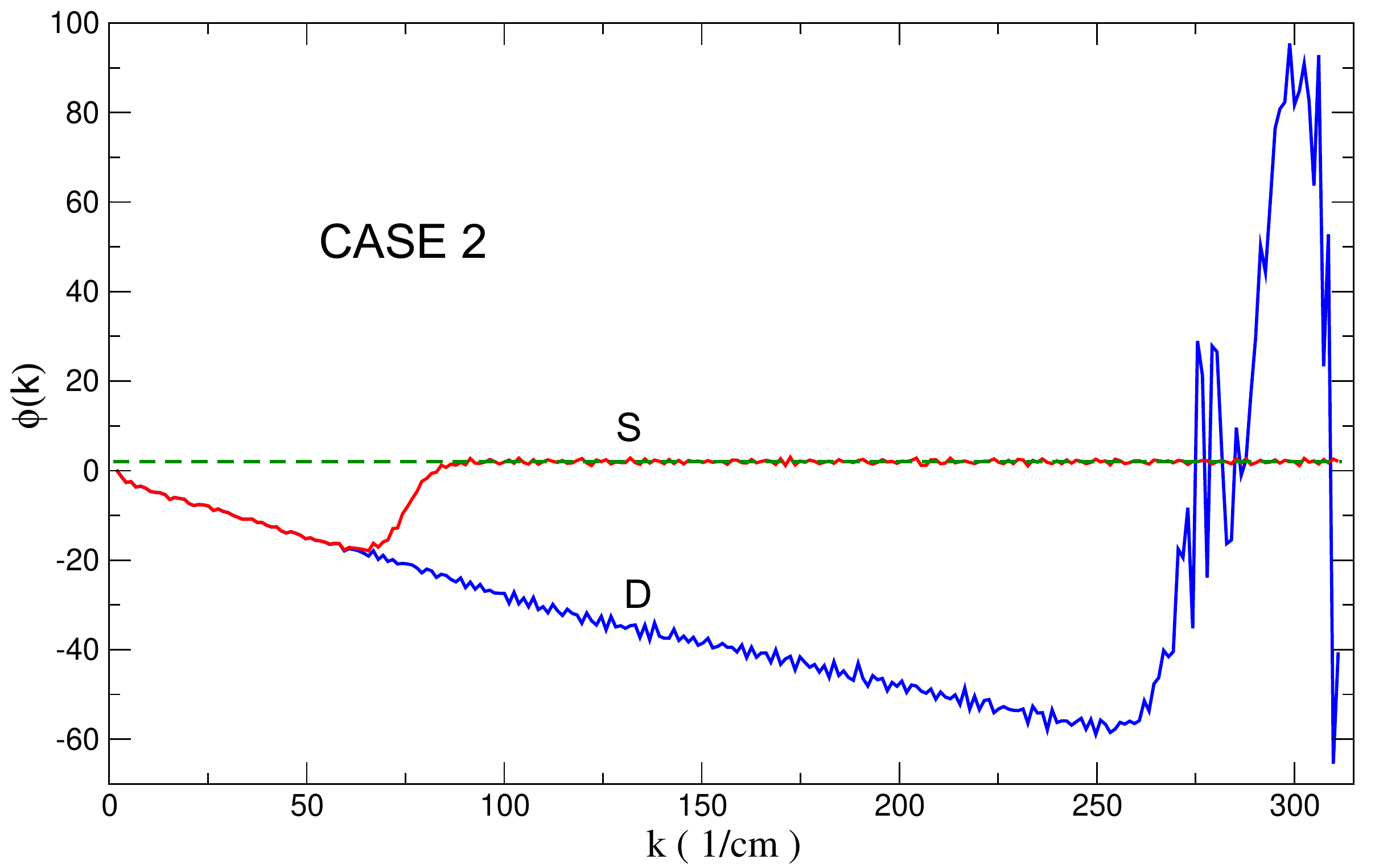}
    \caption{Function $\phi(k)$ for Case~1 (above) and Case~2 (below);  dashed line is $\phi = 2$. Labels Stochastic (S) and Deterministic (D) indicate simulations run with and without thermal fluctuations.}
    \label{fig:PhiTurb}
\end{figure}

It is useful to define \citep{Khurshid2018PRF,Buaria2020PRF}
\begin{align}
\phi( k ) \equiv \frac{d (\ln E(k))}{d (\ln k)}
= \frac{k}{E}~\frac{dE}{dk},
\end{align}
given that, in the absence of thermal fluctuations, the energy spectrum in the dissipation range can be modeled as,
\begin{align}
E(k) = C (k \KolLen)^\alpha \exp( - \beta (k \KolLen)^\gamma ),
\end{align}
in which case we expect $\phi( k ) = \alpha - \beta \gamma (\KolLen k)^\gamma$.
Figure~\ref{fig:PhiTurb} shows that the function $\phi(k)$ approximately is linear for $k \lessapprox 100~\mathrm{cm}^{-1}$ in Case~1 and for $k \lessapprox 50~\mathrm{cm}^{-1}$ in Case~2. Above these wavenumbers for the simulations with thermal fluctuations, the function rises rapidly, plateauing at $\phi(k) \approx 2$, as expected. 
For the deterministic runs the compensated function $\tilde{\phi}(k) \equiv \phi(k)/(\gamma (k \KolLen)^\gamma)$ is approximately constant with $\tilde{\phi} \approx -5.5$ and $-4.7$ for $\gamma \approx 0.85$ and $0.9$ in Cases 1 and 2, respectively.
The deterministic results for $\phi$ and $\tilde{\phi}$ are in reasonable agreement with those reported by \citep{Khurshid2018PRF} and \citep{Buaria2020PRF}, but the runs with thermal noise confirm the original insight of \citep{Betchov1957} that equipartition should occur at wavenumbers near the Kolmogorov scale. 

We also measured the velocity derivative skewness and kurtosis as,
\begin{align}
    \Skew = -\, \frac{\overline{u_{1,1}^3}}{(\overline{u_{1,1}^2})^{3/2}}
    \qquad\mathrm{and}\qquad
    \Kurt = \frac{\overline{u_{1,1}^4}}{(\overline{u_{1,1}^2})^2},
\end{align}
where
\begin{align}
    \overline{{u}_{i,j}^n} = \frac{1}{V} \int d\br ~ (u_{i,j}(\br))^n
    \qquad\mathrm{and}\qquad u_{i,j}(\br) = \frac{\partial u_i(\br)}{\partial r_j}.
\end{align}
For homogeneous, isotropic turbulence one typically has $\Skew \approx \frac13$ to $\frac12$ and $\Kurt \approx 4$. 
At thermodynamic equilibrium we expect $\Skew = 0$ and $\Kurt = 3$ since thermal fluctuations are Gaussian distributed; this was confirmed in simulations with no external forcing (not shown).
In Case 1 (strong forcing) the measured skewness is $\Skew \approx 0.52$ and the kurtosis is $\Kurt \approx 4.5$; in Case 2 (weak forcing) $\Skew \approx 0.40$ and $\Kurt \approx 4.1$. These values did not change significantly when the simulations were repeated with the thermal noise disabled (i.e., deterministic, forced Navier-Stokes). As functions of time, the skewness and kurtosis are smooth but vary significantly. For example, kurtosis in Case 2 was as low as 3.7 and as high as 4.5. 

\begin{figure}
    \centering
    \includegraphics[width=.49\textwidth]{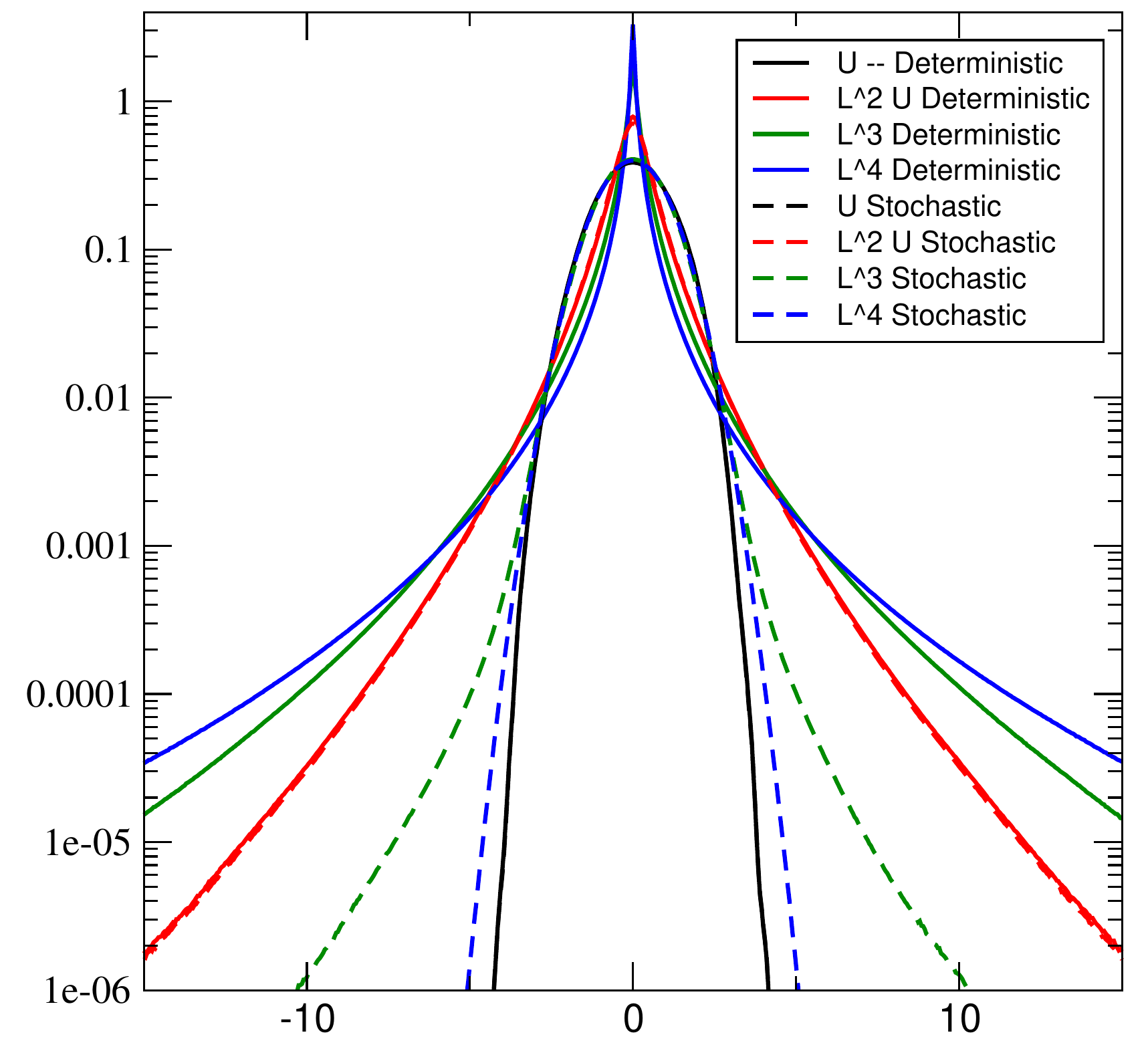}
    \includegraphics[width=.49\textwidth]{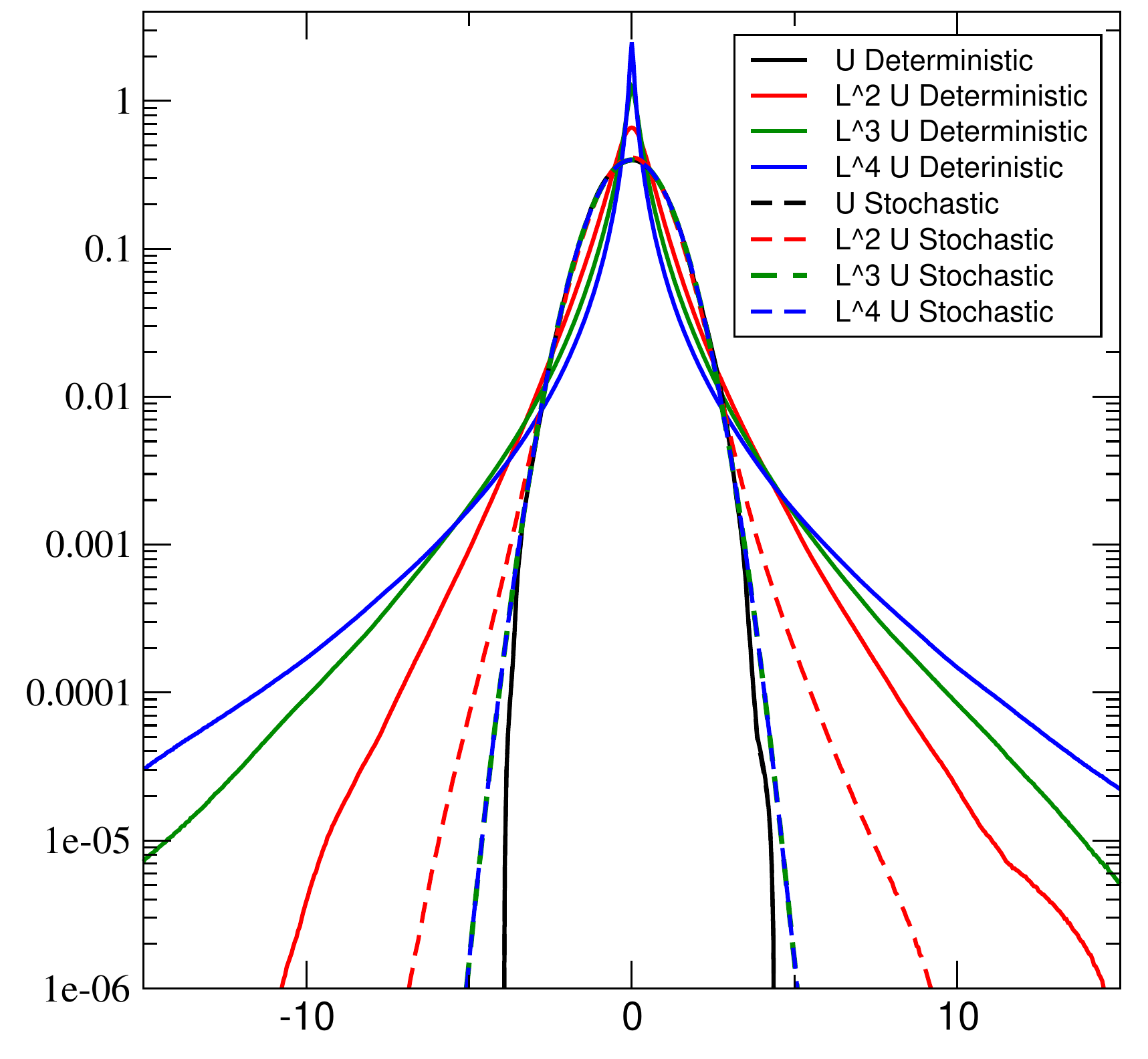}
    
    \caption{PDFs for $\bu$ (black), ${(-\nabla^2})^2 \bu$ (red)
    ${(-\nabla^2})^3 \bu$ (green)
    and ${(-\nabla^2})^4 \bu$ (blue).  In the legend $L$ denotes $-\nabla^2$.
   Left frame is the strong forcing (Case~1). Right frames is the weak forcing (Case~2).
    Note that in both frames solid and dashed black curves overlap and in the right panel the green and blue dashed curves overlap.
    }
    \label{fig:velder_pdfs}
\end{figure}

While the skewness and kurtosis are little affected by thermal noise, \citep{Eyink2021} have argued 
that the extreme intermittency in the far-dissipation range predicted by \citep{Kraichnan1967} is replaced 
by Gaussian thermal equipatition. As in previous numerical studies \citep{Chen1993PRL}, this intermittency
may be diagnosed by probability distribution functions (PDF) for various higher order derivatives of the velocity, with increasing orders probing smaller scales. 
In Figure~\ref{fig:velder_pdfs} we show numerical results for both deterministic and stochastic runs. 
In the absence of thermal fluctuations the PDFs for ${(-\nabla^2})^n \bu,$
with $n=2,3,4$ have approximately exponential tails that grow broader with increasing $n$, as in \citep{Chen1993PRL}. 
In the stochastic simulations we see quite with opposite behavior with the PDF's for ${(-\nabla^2})^n \bu$ 
having suppressed tails for increasing $n$ and becoming nearly Gaussian for $n=4.$
In the absence of large-scale external forcing all three PDFs are Gaussian (not shown), as expected for thermal fluctuations at thermodynamic equilibrium. These results support the prediction of \citep{Eyink2021}.

\section{Concluding remarks}\label{sec:ConclusingRemarks}


It is generally assumed that the large separation between the scale at which turbulent eddies are strongly damped and the molecular mean free path implies that thermal fluctuations are irrelevant in turbulent flows. 
However, statistical mechanics tells us that thermal fluctuations are present at all wavelengths and the fluctuation-dissipation theorem shows their close relation to viscous dissipation. Characterizing the impact of thermal fluctuations on turbulent flow requires the introduction of a dimensionless parameter, $\KolTemp$, related to temperature. For values of $\KolTemp$ representative of flows of interest, the presumption that the viscous dissipation range is well-separated from the thermal range breaks down.
The numerical simulation results in this paper confirm theoretical predictions 
\citep{Betchov1957,Eyink2021}
that the thermal equipartition range in the energy spectrum can dominate in the dissipation range at length scales comparable to the Kolmogorov length. This means that to model sub-Kolmogorov scale turbulence in this regime the conventional incompressible Navier-Stokes equations must be augmented to include thermal fluctuations, as in the Landau-Lifschitz fluctuating Navier-Stokes equations. 
Particle simulations, such as direct simulation Monte Carlo (DSMC), are also useful and preliminary results (Gallis, private communication) also seem to confirm the importance of thermal fluctuations.

While these numerical results agree with theoretical estimates, experimental verification remains crucial but very challenging. Traditional techniques, such as hot-wire anemometry and particle imaging velocimetry, so far lack the resolution and sensitivity to be applicable at the sub-Kolmogorov scales of turbulent flows. Conceptually, it is not even clear that such methods based on continuum fluid descriptions suffice to measure 
molecular velocities coarse-grained over mesoscopic scales. 
It may be more practical to look for indirect evidence of the effects of thermal fluctuations in the dissipation range. For example, high Schmidt/Prandtl-number scalar mixing, droplet and bubble formation, and chemical reactions are all known to depend strongly on sub-Kolmogorov scales but can be equally influenced by thermal noise, 
so that the two effects may compete in determining observed rates and characteristics. 

Finally, in this work we focused on the dissipation range of fully-developed, homogeneous turbulence but we expect thermal fluctuations to be important for other turbulent flow scenarios. For example, they may play an important role in triggering the transition to turbulence. Thermal fluctuations may also be crucial in the generation of unpredictability leading to spontaneous stochasticity. In fact, thermal noise is expected to contribute for any fluid modes that are strongly affected by molecular dissipation, such as the viscous sublayer eddies of wall-bounded turbulence.
For such scenarios and others, fluctuating hydrodynamics provides a powerful theoretical and numerical tool for future work.

\acknowledgements{The authors thank A.~Donev, M.~Gallis, and J.~Goodman for insightful discussions. This work was supported by the U.S.~Department of Energy, Office of Science, Office of Advanced Scientific Computing Research, Applied Mathematics Program under contract No.~DE-AC02-05CH11231 (J.B.,A.N.,A.G). The work of G.E. was supported by the Simons Foundation Targeted Grant in MPS-663054. 
This research used resources of the National Energy Research Scientific Computing Center, a DOE Office of Science User Facility supported 
under Contract No.~DE-AC02-05CH11231.}

\appendix

\section{Nonlinear Fluctuation-Dissipation Relation Proof}


Here, we present
a detailed proof of the nonlinear fluctuation-dissipation relation (FDR)
for a finite-volume discretization of incompressible Navier-Stokes in a periodic space domain.
The proof parallels the proof for the truncated 
``continuum'' fluctuating 
incompressible hydrodynamics using the space Fourier transform to diagonalize the Leray-Hodge projection given by Eyink {\it et al.} \citep{Eyink2021}.
We first briefly review the continuum case.  We then discuss the spatial discretization and extend the FDR to the space-discretized model.  
The elements needed to show the discrete nonlinear FDR are 
that the linearized systems satisfy a discrete FDR, that the inviscid dynamics conserves kinetic energy, and that the inviscid dynamics satisfies a Liouville theorem.  The first two requirements are well known; the key issue is showing the Liouville theorem for the discrete dynamics.

\subsection*{Truncated Continuum Fluctuation-Dissipation Theorem}

The incompressible fluctuating Navier-Stokes equation in the torus domain $\Omega={\mathbb T}^d$
\be \partial_t\bu + (\bu\cdot\grad)\bu = -\grad p + \nu \Delta \bu
-\grad\bdot\tilde{\btau}, \quad \grad\bdot\bu=0. 
\lb{NS} \ee
can be written as a system of stochastic ODE's for the Fourier modes 
\be \hat{\bu}_\bk=\int_\Omega d^3x \ e^{-i\bk\bdot\bx} \bu(\bx) \ee
of the form 
\be
 \partial_t\hat{u}_{\bk,m} + ik_n\left(\delta_{mp}-\frac{k_m k_p}{k^2}\right) 
 \sum_{\bp+\bq=\bk} \hat{u}_{\bp,n}\hat{u}_{\bq,p}  
+\nu k^2\hat{u}_{\bk,m} = \left(\frac{2\nu k_B T}{\rho}\right)^{1/2}ik\, \eta_{\bk,m}(t)
 \lb{FNS-k} \ee
where $\eta_{\bk,m}(t)$ for each wavevector $\bk$ and space component $m$ are
complex white-noise with covariances 
\be \langle \eta_{\bk,m}^*(t)\eta_{\bk',m'}(t')\rangle=V\left(\delta_{mm'}-\frac{k_mk_{m'}}{k^2}\right)\delta_{\bk,\bk'}\delta(t-t'). \ee
Physically this is a ``quasi-continuum'' mesoscopic description valid only at wavenumbers $|\bk|<\Lambda,$ 
some cutoff wavenumber $\ll \lambda_{mfp}^{-1}$ (inverse mean-free path length). For this reason, 
and also to give a precise mathematical meaning to the dynamics, all wavevectors 
in the above dynamical equations are restricted to have magnitudes less than $\Lambda$.

The resulting stochastic dynamics satisfies an exact nonlinear fluctuation-dissipation relation, 
according to which the long-time invariant measure is the Gaussian thermal equilibrium 
distribution 
\be P[\bu] = \frac{1}{Z}\exp\left(-\frac{1}{2k_BT}\sum_{|\bk|<\Lambda}|\hat{\bu}(\bk)|^2\right). \ee
This is a well-known ``folklore'' result, a careful proof of which can be found in Eyink
{\it et al.}~\citep{Eyink2021}.
In fact, this invariant measure is unique because of energy bounds and non-degeneracy of the noise
and is in ``detailed balance'' or time-reversible for the dynamics. The proof in \citep{Eyink2021}
based on the Fokker-Planck equation for the Fourier modes of velocity rests on two key 
results for the inviscid deterministic dynamics given by the truncated Euler equations:
(i) exact conservation of kinetic energy, and (ii) a Liouville Theorem on conservation of 
phase-volume. Conservation of kinetic energy is a consequence of the 
``detailed energy conservation'' for individual triads of Fourier modes, first noted by Onsager \citep{Onsager1949}. 
We comment here briefly on the conservation of phase-volume. 

The Liouville Theorem for truncated Euler was derived by T. D. Lee \citep{Lee1952}, who employed 
the Fourier representation of the dynamics. The statement of this result involves the term 
\be B_{\bk,m}(\hat{\bu},\hat{\bu}^*)= -i k_n\left(\delta_{mp}-\frac{k_m k_p}{k^2}\right) 
 \sum_{\bp+\bq=\bk} \hat{u}_{\bp,n}\hat{u}_{\bq,p} \lb{B-def} \ee 
in Eq.(\ref{FNS-k}). A significant complication, however, is that not all of the Fourier 
modes are independent, because of the reality condition under complex conjugation
\be \hat{\bu}_\bk^*=\hat{\bu}_{-\bk}. \ee 
In his original proof, T.D. Lee used real and imaginary parts of these modes 
for a subset of wavevectors. The proof in \citep{Eyink2021}, Appendix A, instead 
considered the modes whose wavevector lies in the half-set 
\be K^+= \left\{\bk:\ \begin{array}{ll}
                  k_x>0, & \mbox{or} \cr
                  k_y>0 & \mbox{if $k_x=0$, or}\cr 
                  k_z\geq 0 & \mbox{if $k_x=k_y=0$} 
\end{array}\right\} \ee 
and chose $\hat{\bu}_{\bk,m}$ for $\bk\in K^+$ as the independent complex modes. 
This proof used the standard device of treating $\hat{\bu}_{\bk,m}$ and its 
complex conjugate $\hat{\bu}_{\bk,m}^*$ as formally independent variables 
in the calculus of Wirtinger derivatives $\frac{\partial}{\partial \hat{u}_{\bk,m}},$
$\frac{\partial}{\partial \hat{u}_{\bk,m}^*},$ simplifying the original calculations 
of Lee. Here we note that the wavenumbers $\bp,$ $\bq$ which are summed 
over in Eq.(\ref{B-def}) may lie in the complementary set $K^-= -K^+$ and when 
$\bp\in K^-,$ then $\hat{\bu}_\bp$ should be interpreted instead as $\hat{\bu}^*_{-\bp}.$
There is a corresponding equation of motion for the complex-conjugate variables
\be \partial_t\hat{u}_{\bk,m}^* = B_{\bk,m}^*[\hat{\bu},\hat{\bu}^*] -\nu k^2\hat{u}_{\bk,m}^*
-\left(\frac{2\nu k_B T}{\rho}\right)^{1/2}ik\, \eta_{\bk,m}^*(t)\ee 
with $B_{\bk,m}^*[\hat{\bu},\hat{\bu}^*] :=B_{\bk,m}[\hat{\bu},\hat{\bu}^*]^*$ when $\bk\in K^+$ and 
$|\bk|<\Lambda.$ The statement of the Liouville Theorem follows from the easily verified results 
\be \frac{\partial}{\partial \hat{\bu}_\bk}\bdot \bB_\bk[\hat{\bu},\hat{\bu}^*]=-(d-1)i\bk\bdot\hat{\bu}(\bzed), 
\quad \frac{\partial}{\partial \hat{\bu}_\bk^*}\bdot \bB_\bk^*[\hat{\bu},\hat{\bu}^*]=(d-1)i\bk\bdot\hat{\bu}(\bzed)\ee  
in space dimension $d>1.$ In fact, summing over all independent modes then gives
\be \sum_{\bk\in K^+,|\bk|<\Lambda} 
\left(\frac{\partial}{\partial \hat{\bu}_\bk}\bdot \bB_\bk[\hat{\bu},\hat{\bu}^*]
+\frac{\partial}{\partial \hat{\bu}_\bk^*}\bdot \bB_\bk^*[\hat{\bu},\hat{\bu}^*]\right)=0. \ee 

\subsection*{Centered Finite-Volume Space Discretization} 

We now describe the finite-volume space-discretization for fluctuating incompressible Navier-Stokes 
discussed in Usabiaga \etal~\citep{LLNS_Staggered}, Delong \etal~\citep{DFDB} and Nonaka \etal~\citep{Nonaka2015Camcos}.
We note that the discretization is based on incorporating fluctuations into a classical discretization of Navier-Stokes originally introduced by Harlow and Welch \citep{HarWel65}.
For simplicity we consider only $d=2,$ since that 
suffices to illustrate the basic ideas. We shall also consider only the periodic domain 
${\mathbb T}^2:={\mathbb R}^2/(L_x{\mathbb Z}\times L_y{\mathbb Z})$ and consider a spatial 
discretization $(x_i,y_j)=(i\Delta x,j\Delta y)$ with $0\leq i<N_x,$ $0\leq j<N_y$ and 
$L_x=N_x\Delta x,$ $L_y=N_y\Delta y.$ In this scheme, scalar fields like pressure $p$
live on cell centers at lattice sites $(x_i,y_j)$, denoted $p_{i,j}.$ Vector components
live on cell faces displaced in the corresponding directions,  so that $x$-component of 
velocity is $u_{i+\half,j}$ and $y$-component of velocity is $v_{i,j+\half}$.
The spatially-discretized equations of motion (but continuous in time) have the form,
ignoring for the moment stochastic terms: 
\bea 
\dot{u}_{i+\half,j}&=& -\grad\bdot(\bv u)_{i+\half,j} -(\nabla_x p)_{i+\half,j} -\nu(\Delta u)_{i+\half,j}\cr
\dot{v}_{i,j+\half}&=& -\grad\bdot(\bv v)_{i,j+\half} -(\nabla_y p)_{i,j+\half} -\nu(\Delta v)_{i,j+\half}
\label{eq:first}
\eea
where all gradients denote centered-differences, for example,
\be (\nabla_x p)_{i+\half,j}=\frac{p_{i+1,j}-p_{i,j}}{\Delta x}, \quad 
(\nabla_y p)_{i,j+\half}=\frac{p_{i,j+1}-p_{i,j}}{\Delta y} \ee 
and $\Delta$ is the standard 5-point laplacian. The nonlinear terms are 
calculated on interpolated lattice sites by averaging adjacent values. Thus,
\bea 
\grad\bdot(\bv u)_{i+\half,j}&=& \frac{1}{\Delta x}\left[\overline{u}_{i+1,j}^2-\overline{u}_{i,j}^2\right]
+\frac{1}{\Delta y}\left[\overline{u}_{i+\half,j+\half}\overline{v}_{i+\half,j+\half}
-\overline{u}_{i+\half,j-\half}\overline{v}_{i+\half,j-\half}\right] \cr
\grad\bdot(\bv v)_{i,j+\half}&=& \frac{1}{\Delta x}\left[\overline{u}_{i+\half,j+\half}\overline{v}_{i+\half,j+\half}
-\overline{u}_{i-\half,j+\half}\overline{v}_{i-\half,j+\half}\right]
+\frac{1}{\Delta y}\left[\overline{v}_{i,j+1}^2-\overline{v}_{i,j}^2\right]
\eea
where, for example, 
\bea 
&& \overline{u}_{i,j}=\frac{u_{i+\half,j}+u_{i-\half,j}}{2}, \quad 
\overline{v}_{i,j}=\frac{v_{i,j+\half}+v_{i,j-\half}}{2}, \cr 
&& \overline{u}_{i+\half,j+\half}=\frac{u_{i+\half,j+1}+u_{i+\half,j}}{2}, \quad 
\overline{v}_{i+\half,j+\half}=\frac{v_{i+1,j+\half}+v_{i,j+\half}}{2}, \quad\mathrm{etc.} 
\eea 
The velocity field satisfies the discrete incompressibility condition
\be (\nabla_x u)_{ij}+(\nabla_y v)_{ij}= \frac{u_{i+\half,j}-u_{i-\half,j}}{\Delta x}
+ \frac{v_{i,j+\half}-v_{i,j-\half}}{\Delta y} =0, \ee 
which implies a Poisson equation to determine the pressure:
\be (-\nabla^2 p)_{ij} =\frac{1}{\Delta x}\left[\grad\bdot(\bv u)_{i+\half,j}-\grad\bdot(\bv u)_{i-\half,j}\right]
+\frac{1}{\Delta x}\left[\grad\bdot(\bv v)_{i,j+\half}-\grad\bdot(\bv v)_{i,j-\half}\right]
\label{eq:last}
\ee 

In order to prove the Liouville theorem for the space-discretized dynamics, it is convenient,
just as for the continuum case, to use Fourier modes. We introduce the discrete Fourier transforms 
\be a_{k,\ell} = \frac{1}{N} \sum_{i,j} e^{-i(ki\Delta x+\ell j\Delta y)}u_{i+\half,j}, \quad
b_{k,\ell} = \frac{1}{N} \sum_{i,j} e^{-i(ki\Delta x+\ell j\Delta y)}v _{i,j+\half}, 
\ee 
\be q_{k,\ell} = \frac{1}{N} \sum_{i,j} e^{-i(ki\Delta x+\ell j\Delta y)}p_{i,j}, \ee
with $N=N_xN_y$ and with $k\in 2\pi {\mathbb Z}_{N_x}/L_x,$ $\ell\in 3\pi {\mathbb Z}_{N_y}/L_y$
In that case, reality of the basic variables 
$u_{i+\half,j},$ $v _{i,j+\half},$ implies that relations $a_{k,\ell}^*=a_{-k,-\ell},$ $b_{k,\ell}^*=b_{-k,-\ell}$ hold.  
The consequence is that not all of these variables are independent and, 
in the proof of the Liouville Theorem, we must select an independent subset. 

It is convenient here for bookkeeping purposes to label these modes as $a_{\alpha,\beta},$ $b_{\alpha,\beta}$
using the integers $\alpha\in {\mathbb Z}_{N_x},$  $\beta\in {\mathbb Z}_{N_y},$ 
where $k_\alpha=2\pi\alpha/L_x,$ $\ell_\beta=2\pi\beta/L_y.$ In that case we may choose
$0\leq \alpha<N_x,$ $0\leq \beta<N_y$ as representative values. Consider first the case 
with $N_x,$ $N_y$ both odd. In this case the reality conditions become 
\begin{eqnarray} 
\mbox{$\alpha\neq 0$:\ $a_{\alpha,\beta}^*=a_{N_x-\alpha,N_y-\beta}$} &\Longrightarrow & \mbox{$\alpha$ can be restricted 
to $1\leq\alpha \leq \frac{N_x-1}{2}$} \cr 
\mbox{$\beta\neq 0$:\ $a_{0,\beta}^*=a_{0,N_y-\beta}$} & \Longrightarrow & \mbox{$\beta$ can be restricted 
to $1\leq\beta \leq \frac{N_y-1}{2}$}\cr 
a_{0,0}^*=a_{0,0} &\Longrightarrow & \mbox{$a_{0,0}$ is real}
\end{eqnarray}
and similarly for $b_{\alpha,\beta}.$ We may thus take as independent variables the complex 
quantities $a_{\alpha,\beta}$ for $1\leq\alpha \leq \frac{N_x-1}{2},$ $0\leq \beta<N_y$
and $a_{0,\beta}$ for $1\leq\beta \leq \frac{N_y-1}{2}$ and the single real variable $a_{0,0},$
and similarly for $b_{\alpha,\beta}.$ On the other hand, with $N_x,$ $N_y$ both even,
the reality conditions become 
\begin{eqnarray} 
\mbox{$\alpha\neq 0,\frac{N_x}{2}$:\ 
$a_{\alpha,\beta}^*=a_{N_x-\alpha,N_y-\beta}$} &\Longrightarrow & \mbox{$\alpha$ can be restricted 
to $1\leq\alpha \leq \frac{N_x-2}{2}$} \cr 
\mbox{$\beta\neq 0,\frac{N_y}{2}$:\ $a_{0,\beta}^*=a_{0,N_y-\beta}$, $a_{\frac{N_x}{2},\beta}^*=a_{\frac{N_x}{2},N_y-\beta}$} & \Longrightarrow & \mbox{$\beta$ can be restricted 
to $1\leq\beta \leq \frac{N_y-1}{2}$}\cr 
a_{0,0}^*=a_{0,0},a_{0,\frac{N_y}{2}}^*=a_{0,\frac{N_y}{2}},a_{\frac{N_x}{2},0}^*
=a_{\frac{N_x}{2},0},a_{\frac{N_x}{2},\frac{N_y}{2}}^*=a_{\frac{N_x}{2},\frac{N_y}{2}} &\Longrightarrow & 
\mbox{$a_{0,0},a_{0,\frac{N_y}{2}},a_{\frac{N_x}{2},0},a_{\frac{N_x}{2},\frac{N_y}{2}}$ are real}\cr
&&
\end{eqnarray} 
We may thus take as independent variables the complex 
quantities $a_{\alpha,\beta}$ for $1\leq\alpha \leq \frac{N_x-2}{2},$ $0\leq \beta<N_y$
and $a_{0,\beta},$ $a_{\frac{N_x}{2},\beta}$ for $1\leq\beta \leq \frac{N_y-2}{2},$ and the 
four real variables $a_{0,0},a_{0,\frac{N_y}{2}},a_{\frac{N_x}{2},0},a_{\frac{N_x}{2},\frac{N_y}{2}},$
and similarly for $b_{\alpha,\beta}.$ The cases with one of $N_x,$ $N_y$ odd and the other even can be treated
likewise. 

The inverse relations hold 
\be  u_{i+\half,j}= \sum_{k,\ell} e^{i(ki\Delta x+\ell j\Delta y)}a_{k,\ell}, \quad
 v _{i,j+\half}= \sum_{k,\ell} e^{i(ki\Delta x+\ell j\Delta y)} b_{k,\ell} 
\ee 
\be  p_{i,j}= \sum_{i,j} e^{i(ki\Delta x+\ell j\Delta y)}q_{k,\ell}.  \ee 
We then find that spatial derivatives are given by 
\be  (\nabla_x u)_{i,j}= \sum_{i,j} e^{i(ki\Delta x+\ell j\Delta y)} ik^- a_{k,\ell}, \quad
 (\nabla_y v)_{i,j}= \sum_{i,j} e^{i(ki\Delta x+\ell j\Delta y)} i\ell^- b_{k,\ell} 
\ee 
with 
\be k^-:= \frac{1}{i\Delta x}(1-e^{-ik\Delta x}), \quad \ell^-:= \frac{1}{i\Delta y}(1-e^{-i\ell\Delta y})\ee 
and the complex conjugates $k^+=(k^-)^*,$ $k^+=(k^-)^*$ given by 
\be k^+:= \frac{1}{i\Delta x}(e^{ik\Delta x}-1), \quad \ell^+:= \frac{1}{i\Delta y}(e^{i\ell\Delta y}-1)\ee 
Similarly, the discrete Laplacian Fourier transforms as 
\be -\widehat{(\nabla^2 p)}_{k,\ell} =(|k^+|^2+|\ell^+|^2)q_{k,\ell} = 
\left[\frac{4}{(\Delta x)^2} \sin^2\left(\frac{k\Delta x}{2}\right)
+\frac{4}{(\Delta y)^2} \sin^2\left(\frac{\ell\Delta y}{2}\right)\right] q_{k,\ell} \ee 
Lastly, we note that averaged fields can be Fourier analyzed as well, for example 
\be  \overline{u}_{i+\half,j+\half}= \sum_{k,\ell} e^{i(ki\Delta x+\ell j\Delta y)} \overline{a}_{k,\ell}^{(+y)}, \quad
 \overline{v} _{i+\half,j+\half}= \sum_{k,\ell} e^{i(ki\Delta x+\ell j\Delta y)} \overline{b}_{k,\ell}^{(+x)}  
\ee  
with 
\be \overline{a}_{k,\ell}^{(\pm y)}:= \frac{1}{2}(1+e^{\pm ik \Delta y})a_{k,\ell}, 
\quad  \overline{b}_{k,\ell}^{(\pm x)}:= \frac{1}{2}(1+e^{\pm ik \Delta x})b_{k,\ell} \ee
and similarly for other fields. 

With these definitions we note that a straightforward but tedious calculation gives the 
deterministic dynamics of Fourier modes as: 
\bea
\dot{a}_{k,\ell} &=& -ik^+\sum_{k',\ell'} \overline{a}_{k',\ell'}^{(-x)} \overline{a}_{k-k',\ell-\ell'}^{(-x)} 
-i\ell^-\sum_{k',\ell'} \overline{a}_{k',\ell'}^{(+y)} \overline{b}_{k-k',\ell-\ell'}^{(+x)} -ik^+q_{k,\ell} 
-\nu (|k^+|^2+|\ell^+|^2) a_{k,\ell} 
\cr 
\dot{b}_{k,\ell} &=& -ik^-\sum_{k',\ell'} \overline{a}_{k',\ell'}^{(+y)} \overline{b}_{k-k',\ell-\ell'}^{(+x)} 
-i\ell^+\sum_{k',\ell'} \overline{b}_{k',\ell'}^{(-y)} \overline{b}_{k-k',\ell-\ell'}^{(-y)} -i\ell^+q_{k,\ell} 
-\nu (|k^+|^2+|\ell^+|^2) b_{k,\ell} 
\eea
We next prove for this dynamics the two essential ingredients needed for the nonlinear FDR, namely:
(i) exact conservation of kinetic energy and (ii) the Liouville Theorem on 
conservation of phase volume. We begin with the latter. 

\subsection*{Discrete Liouville Theorem}

We introduce the following notation for the inviscid part of the dynamics
\bea
{A}_{k,\ell} &=& -ik^+\sum_{k',\ell'} \overline{a}_{k',\ell'}^{(-x)} \overline{a}_{k-k',\ell-\ell'}^{(-x)} 
-i\ell^-\sum_{k',\ell'} \overline{a}_{k',\ell'}^{(+y)} \overline{b}_{k-k',\ell-\ell'}^{(+x)} -ik^+q_{k,\ell} \cr 
{B}_{k,\ell} &=& -ik^-\sum_{k',\ell'} \overline{a}_{k',\ell'}^{(+y)} \overline{b}_{k-k',\ell-\ell'}^{(+x)} 
-i\ell^+\sum_{k',\ell'} \overline{b}_{k',\ell'}^{(-y)} \overline{b}_{k-k',\ell-\ell'}^{(-y)} -i\ell^+q_{k,\ell} 
\eea
and state our main result: 

\begin{Prop}
The formula holds
\be \frac{\partial A_{k,\ell}}{\partial a_{k,\ell}} + \frac{\partial B_{k,\ell}}{\partial b_{k,\ell}}
= -i \frac{\sin(k\Delta x)}{\Delta x} a_{0,0} -i \frac{\sin(\ell\Delta y)}{\Delta y} b_{0,0} 
\ee
and thus
\be 
\sum_{\mbox{{\small complex modes} }(k,\ell)} \left(\frac{\partial A_{k,\ell}}{\partial a_{k,\ell}} + \frac{\partial B_{k,\ell}}{\partial b_{k,\ell}}
+\frac{\partial A_{k,\ell}^*}{\partial a_{k,\ell}^*} + \frac{\partial B_{k,\ell}^*}{\partial b_{k,\ell}^*}\right) 
+\sum_{\mbox{{\small real modes} } (k,\ell)} 
\left(\frac{\partial A_{k,\ell}}{\partial a_{k,\ell}} + \frac{\partial B_{k,\ell}}{\partial b_{k,\ell}}\right) =0
\ee 
Note that these are the discrete analogues of the continuum results (10),(11) for $d=2.$
\end{Prop} 
{\it Proof:} Note that the advective part of the dynamics is represented by 
\bea
{A}_{k,\ell}^{adv} &=& -ik^+\sum_{k',\ell'} \overline{a}_{k',\ell'}^{(-x)} \overline{a}_{k-k',\ell-\ell'}^{(-x)} 
-i\ell^-\sum_{k',\ell'} \overline{a}_{k',\ell'}^{(+y)} \overline{b}_{k-k',\ell-\ell'}^{(+x)} \cr 
{B}_{k,\ell}^{adv} &=& -ik^-\sum_{k',\ell'} \overline{a}_{k',\ell'}^{(+y)} \overline{b}_{k-k',\ell-\ell'}^{(+x)} 
-i\ell^+\sum_{k',\ell'} \overline{b}_{k',\ell'}^{(-y)} \overline{b}_{k-k',\ell-\ell'}^{(-y)} 
\eea
and the Poisson equation for the pressure in Fourier representation becomes
\be q_{k,\ell}= - \frac{ik^- A_{k,\ell}^{adv}+i\ell^-B_{k,\ell}^{adv}}{|k^+|^2+|\ell^+|^2}.  
\ee
Thus, the inviscid dynamics can be represented via a discrete Leray projection as 
\bea
A_{k,\ell} &=& \frac{|\ell^+|^2}{|k^+|^2+|\ell^+|^2} A_{k,\ell}^{adv}-\frac{k^+\ell^-}{|k^+|^2+|\ell^+|^2}B_{k,\ell}^{adv} \cr 
B_{k,\ell} &=& -\frac{\ell^+ k^-}{|k^+|^2+|\ell^+|^2}A_{k,\ell}^{adv} +\frac{|k^+|^2}{|k^+|^2+|\ell^+|^2} B_{k,\ell}^{adv}
\eea
The following straightforward derivatives 
\bea
\frac{\partial A_{k,\ell}^{adv}}{\partial a_{k,\ell}} &=& -2i\frac{\sin(k\Delta x)}{\Delta x} a_{0,0} 
-i \frac{\sin(\ell\Delta y)}{\Delta y}b_{0,0}, \quad 
\frac{\partial A_{k,\ell}^{adv}}{\partial b_{k,\ell}} \ =\ -i\ell^- \frac{1}{2}(1+e^{ik\Delta x})a_{0,0} \cr  
\frac{\partial B_{k,\ell}^{adv}}{\partial b_{k,\ell}} &=& -i \frac{\sin(k\Delta x)}{\Delta x}a_{0,0}
-2i\frac{\sin(\ell\Delta y)}{\Delta y} b_{0,0}, \quad 
\frac{\partial B_{k,\ell}^{adv}}{\partial a_{k,\ell}} \ =\ -ik^- \frac{1}{2}(1+e^{i\ell\Delta y})b_{0,0} 
\eea
together with (36) yields the result (32). 

Finally, we note that the expression in (32) for complex modes is pure imaginary 
and thus cancels in (33) with the contribution from the complex conjugate. On the other 
hand, for real modes the expressions in (32) vanish individually because $k\Delta x,$ 
$\ell\Delta y$ are equal either to 0 or $\pi.$ \hfill $\Box$

\subsection*{Discrete Energy Conservation}

It is well know that the discretization Eqs. (\ref{eq:first})-(\ref{eq:last})
for the inviscid case $\nu=0$ in periodic boundary 
conditions exactly conserves the discrete kinetic energy (per mass):
\be H= \half\sum_{ij} (u_{i+\half,j}^2+v_{i,j+\half}^2). \ee
This result follows from the skew-adjoint property of the advection discretization applied to discretely divergence free fields and orthogonality of discrete gradients with discretely divergence free fields.  See, for example,
Delong \etal~\citep{DFDB}, Usabiaga \etal~\citep{LLNS_Staggered}.
We summarize the argument below for completeness.

To prove the first statement, we note using $(u_{i+\half,j}-u_{i-\half,j})\overline{u}_{i,j}=
\half(u_{i+\half,j}^2-u_{i-\half,j}^2)$ that 
\bea \frac{1}{\Delta x} \sum_{ij} u_{i+\half,j}(\overline{u}_{i+1,j}^2-\overline{u}_{i,j}^2) 
&=& -\frac{1}{\Delta x} \sum_{ij} (u_{i+\half,j}-u_{i-\half,j})\overline{u}_{i,j}^2\cr 
&=& -\frac{1}{\Delta x} \sum_{ij} \half(u_{i+\half,j}^2-u_{i-\half,j}^2)\overline{u}_{i,j} \cr
&=& \frac{1}{\Delta x} \sum_{ij} \half u_{i+\half,j}^2(\overline{u}_{i+1,j}-\overline{u}_{i,j})  
\eea 
An exactly analogous calculation shows that 
\be \frac{1}{\Delta y} \sum_{ij} u_{i+\half,j}(\overline{u}_{i+\half,j+\half}\overline{v}_{i+\half,j+\half} 
-\overline{u}_{i+\half,j-\half}\overline{v}_{i+\half,j-\half}) = 
\frac{1}{\Delta y} \sum_{ij} \half u_{i+\half,j}^2(\overline{v}_{i+\half,j+\half} 
-\overline{v}_{i+\half,j-\half}). 
\ee 
Adding these results gives 
\be \sum_{ij} u_{i+\half,j} [\grad\bdot(\bu u)]_{i+\half,j} =0 \ee
since 
\be \frac{1}{\Delta x} (\overline{u}_{i+1,j}-\overline{u}_{i,j}) +
\frac{1}{\Delta y} (\overline{v}_{i+\half,j+\half} -\overline{v}_{i+\half,j-\half})=0 \ee 
is implied by discrete incompressibility.  This shows conservation of 
$\half\sum_{ij} u_{i+\half,j}^2$ by discretized advection and conservation of 
$\half\sum_{ij} v_{i,j+\half}^2$ follows by an identical argument. 

The conservation of total energy by the pressure gradient is more direct, and 
follows from discrete incompressibility and the fact that the finite-volume discretizations
of the gradient operator $\bG$ and divergence operator $\bD$ satisfy $\bG^*=-\bD.$

\subsection*{Proof of the Nonlinear FDR}

To complete the proof of the nonlinear FDR, we note that Usabiaga \etal~\citep{LLNS_Staggered}
added noise to the discretized Stokes equation so that 
$P=(1/Z)\exp(-H/k_BT)$ is the exact stationary measure of this linear stochastic 
dynamics. This was guaranteed by adding the noise in the form 
\be \partial_t\bv = -\bG p+\nu \bL \bv
+\tilde{\bF} \ee 
where $\bL=\bD\bG$ is the discrete Laplacian, where for $\Delta V=\Delta x\Delta y$
\be \tilde{\bF} = \bD\left( \sqrt{\frac{2\nu k_BT}{\rho \Delta V}}\bW\right) \ee
and where $\bW$ is a space-discretized set of temporal white noises living on the 
faces of the shifted velocity grids, e.g.~in 2D consisting of two independent 
white noises $W^{(x)}_{i,j},$ $W^{(y)}_{i,j},$ at the cell centers and 
another two $W^{(x)}_{i+\half,j+\half},$ $W^{(y)}_{i+\half,j+\half}$ at the 
corner points/nodes. This result is easily verified by taking discrete Fourier transforms. 
Adding this same noise into the discretized nonlinear equations (12), the invariant measure is 
preserved. Indeed, because of energy conservation and Liouville Theorem, the gaussian Gibbs measure 
is also invariant for the inviscid deterministic dynamics. See Eyink \etal~(2021) for more details.

\bibliography{TurbFHD}

\end{document}